\documentclass[12pt]{article}
\usepackage{a4wide,epsfig,amsmath,amssymb,cite,scalefnt}

\def\sic{supersymmetric}
\def\semi{;\hfil\break}
\newcommand{\api}{\frac{\alpha_s}{\pi}}

\newcommand{\epscalar}{$\varepsilon$-scalar}
\newcommand{\reference}[1]{Ref.~\cite{#1}}
\newcommand{\abbrev}{\scalefont{.9}}
\newcommand{\drbar}{$\overline{\mbox{\abbrev DR}}$}
\newcommand{\msbar}{$\overline{\mbox{\abbrev MS}}$}
\newcommand{\asDRbar}{\alpha_s^{\overline{\rm DR}}}
\newcommand{\asMSbar}{\alpha_s^{\overline{\rm MS}}}
\newcommand{\betaDRbar}{\beta^{\overline{\rm DR}}}
\newcommand{\betaMSbar}{\beta^{\overline{\rm MS}}}
\newcommand{\gammaDRbar}{\gamma^{\overline{\rm DR}}}
\newcommand{\gammaMSbar}{\gamma^{\overline{\rm MS}}}
\newcommand{\ZsDRbar}{Z_s^{\overline{\rm DR}}}

\newcommand{\ZmDRbar}{Z_m^{\overline{\rm DR}}}

\newcommand{\apiDR}{\frac{\asDRbar}{\pi}}
\newcommand{\apiMS}{\frac{\asMSbar}{\pi}}
\newcommand{\aepi}{\frac{\alpha_e}{\pi}}

\newcommand{\qcd}{{\abbrev QCD}}
\newcommand{\susy}{{\abbrev SUSY}}
\newcommand{\mssm}{{\abbrev MSSM}}
\newcommand{\dreg}{{\abbrev DREG}}
\newcommand{\dred}{{\abbrev DRED}}
\newcommand{\nsvz}{{\abbrev NSVZ}}

\newcommand{\mDRbar}{m^{\overline{\rm DR}}}
\newcommand{\mMSbar}{m^{\overline{\rm MS}}}

\begin{document}

\title{\vskip-3cm{\baselineskip14pt
    \begin{flushleft}
      \normalsize SFB/CPP-06-48 \\
      \normalsize TTP/06-27  \\
      \normalsize WUB/06-05 \\
      \normalsize LTH723  \\
      \normalsize NSF-KITP-06-86
  \end{flushleft}}
  \vskip1.5cm
  Four-loop $\beta$ function and mass anomalous dimension in 
  Dimensional Reduction
}
\author{\small R.V. Harlander$^{(a)}$, 
  D.R.T. Jones$^{(b)}$,
  P. Kant$^{(c)}$,
  L. Mihaila$^{(c)}$, M. Steinhauser$^{(c)}$\\[1em]
  {\small\it (a) Fachbereich C, Theoretische Physik,
    Universit{\"a}t Wuppertal,}\\
  {\small\it 42097 Wuppertal, Germany}\\
  {\small\it (b) Department of Mathematical Sciences, 
    University of Liverpool,}\\
  {\small\it Liverpool L69 3BX, UK}\\
  {\small\it (c) Institut f{\"u}r Theoretische Teilchenphysik,
    Universit{\"a}t Karlsruhe,}\\
  {\small\it 76128 Karlsruhe, Germany}\\
}

\date{}

\maketitle

\thispagestyle{empty}

\begin{abstract}
  Within the framework of \qcd{} we compute renormalization constants
  for the strong coupling and the quark masses to four-loop order. We
  apply the \drbar{} scheme and put special emphasis on the additional
  couplings which have to be taken into account. This concerns the
  \epscalar{}--quark Yukawa coupling as well as the vertex containing four
  \epscalar{}s.  For a supersymmetric Yang Mills theory, we find, in
  contrast to a previous claim, that the evanescent Yukawa coupling 
  equals the strong coupling
  constant through three loops as required by supersymmetry.
\medskip


\end{abstract}

\section{\label{sec::intro}Introduction}

Dimensional regularisation~\cite{'tHooft:1972fi, Bollini:1972ui},  
(\dreg) is a remarkably elegant procedure which completely dominates the
radiative corrections industry associated with the standard model.  
Advocates of alternative regularisation methods rarely proceed  beyond
one  loop (or exceptionally two).  The fundamental reason for the
\dreg{} hegemony is that (with little increase in calculational
difficulty) it preserves  gauge invariance; that is to say,  when the
effective action is separated into a finite part (which is  retained)
and an ``infinite'' part (or more precisely, a part which tends to
infinity in the limit that $D = 4 -2\epsilon \to 4$) then  the finite
effective action  satisfies  the Ward identities of the gauge symmetry,
without the necessity of introducing additional  finite local
counter-terms.

\dreg{} is, however, less well-suited for supersymmetric theories
because invariance of a given  action  with respect to supersymmetric
transformations only holds in general for  specific values of the
space-time dimension $D$. An elegant attempt to modify \dreg{} so as to
render it  compatible with supersymmetry (\susy{}) was  made by
Siegel~\cite{siegel}. The essential difference between Siegel's  method  
(\dred\footnote{Dimensional reduction in combination 
with modified minimal subtraction is often known as \drbar; we will 
use the terms \dred{} and \drbar{} interchangeably.}) 
and \dreg{} is that the continuation from $4$ to $D$ dimensions  is 
made by {\it compactification\/}, or {\it dimensional reduction}.  Thus
while the momentum (or space-time) integrals are $D$-dimensional in the
usual way, the number of field components remains unchanged and
consequently  \susy{} is undisturbed. (A pedagogical introduction
to \dred{} was given by Capper et al~\cite{Capper}.)

As pointed out by Siegel himself,~\cite{siegelb} there remain
potential  problems with \dred. One manifestation of this
was demonstrated in \reference{Avdeev:1981vf}, where  it was shown that
the variation $\delta S$ of the action of a pure (no chiral matter)
supersymmetric gauge theory is nonzero even with \dred.
If $\delta S$ gives a nonzero result when inserted in a Greens function this
creates an apparent violation of supersymmetric Ward identities.
With \dreg{} this happens at one loop, but with \dred{} all explicit
calculations to date have found zero for such insertions. For discussion see
Refs.~\cite{jjs,ds}.

We turn now to the application of \dred{} to non-supersymmetric
theories. That \dred{} is a viable alternative to \dreg{} in the
non-\sic\ case   was claimed early on~\cite{Capper}.  Subsequently it
was adopted occasionally,  motivated, for example, by the fact that Dirac
matrix algebra is easier  in four dimensions and in particular by the
desire to use  Fierz identities. One must, however,  be very careful in
applying \dred{} to non-supersymmetric theories because of the existence
of  {\it evanescent couplings}. These were first  described~\cite{Tim}\
in 1979,  and  later independently by van Damme and 't
Hooft~\cite{hvand}.  After dimensional reduction the four-dimensional
vector field can be decomposed into a $D$-dimensional vector field and a
$2\epsilon$-dimensional which transforms  under gauge transformations as
a scalar and is hence known  as an \epscalar{}. Couplings
involving the \epscalar{}  are called  evanescent    couplings; in
a non-\sic\ theory  they renormalise in a manner different
 from the ``real'' couplings with which we may be tempted to associate
them. It has been conclusively demonstrated~\cite{Jack:1993ws, dredb}
that there  exists  a   set of transformations  whereby the
$\beta$--functions of a particular  theory (calculated using \dred) may 
be related to the $\beta$--functions of the same theory (calculated
using \dreg) by  means of coupling constant reparametrisation. The
evanescent couplings play a crucial role in this analysis, but
in the literature on non-\sic\ \dred{} their significance is often ignored,
and there have been few calculations  which explicitly take them into
account.  In a recent paper~\cite{Harlander:2006rj}, four of us
described in detail the  evanescent coupling structure of \qcd{} and 
calculated the gauge $\beta$-function,  $\beta$, and the mass anomalous
dimension $\gamma_m$ through three loops.  We found that at three loops
$\beta$ depends on the \epscalar{}  Yukawa coupling $g_e$, while
$\gamma_m$ depends  on both $g_e$ and the \epscalar{} quartic self
couplings, $\lambda_r$. In this paper we extend these calculations to
the four-loop level, when  $\beta$ also depends on $\lambda_r$. These results 
bring the precision of our knowledge of these quantities in \dred{} up to 
that in \dreg{}~\cite{vanRitbergen:1997va,Chetyrkin:1997dh,
Vermaseren:1997fq,Czakon:2004bu}.

\section{\label{sec::evan}Evanescent couplings}

The technical framework of our calculation is described in detail in
\reference{Harlander:2006rj}. Let us at this point emphasise once more
the role of the evanescent couplings and in particular elaborate on the
$\varepsilon\varepsilon\varepsilon\varepsilon$ vertex.  The part of the
Lagrange density describing the latter is given by
\begin{eqnarray}
  {\cal L} &=& \ldots 
  -\frac{1}{4}\sum_{r=1}^R\lambda_r
  H_r^{abcd}
  \varepsilon_\sigma^a \varepsilon_{\sigma^\prime}^c 
  \varepsilon_\sigma^b \varepsilon_{\sigma^\prime}^d
  + \ldots
  \,,
\label{eq::lagrange}
\end{eqnarray}
where the $\varepsilon_\sigma^a$ are the \epscalar{} fields, and
$\sigma$, $\sigma^\prime$ are $2 \varepsilon$-dimensional indices. For
the gauge group $SU(N)$ the dimensionality $R$ of the basis for
rank-four tensors is given by $R=3$ for $SU(2)$, $R=8$ for $SU(3)$ and
$R=9$ for $SU(N)$, $N\geq 4$; for tensors $H_r^{abcd}$ symmetric with
respect to $(a,b)$ and $(c,d)$ exchange these numbers become $R = 2$, $R
= 3$ and $R = 4$ respectively~\cite{Jack:1993ws}.  We will restrict
ourselves to $SU(3)$; our basis choice reads
\begin{eqnarray}
  H^{abcd}_1 &=& \frac{1}{2}\left (f^{ace} f^{bde}  + f^{ade} f^{bce}\right) 
    \,,\nonumber\\
  H^{abcd}_2 &=& \delta^{ab} \delta^{cd} 
  + \delta^{ac} \delta^{bd}  + \delta^{ad} \delta^{bc} 
  \,,\nonumber\\
  H^{abcd}_3 &=&  \frac{1}{2}\left(
  \delta^{ac} \delta^{bd} + \delta^{ad} \delta^{bc}\right)
  - \delta^{ab} \delta^{cd}
  \,.
\end{eqnarray}
Note that dimensional reduction of the original action yields 
$\lambda_1 = g^2$, $\lambda_2 = \lambda_3 = 0$, but, as we 
have already emphasised, this situation is not preserved by renormalisation. 

Once the tensors $H_r^{abcd}$ are chosen, the Feynman rules are fixed
in a unique way.
It is straightforward to relate the result obtained from a different
choice of the $H_r^{abcd}$ to each other. For example, for
\begin{eqnarray}
  \tilde H^{abcd}_1 &=& \frac{1}{2}  \delta^{ab} \delta^{cd} 
  \,,\nonumber\\
  \tilde H^{abcd}_2 &=&  \frac{1}{2} \left(\delta^{ac} \delta^{bd} +
  \delta^{ad} \delta^{bc} \right)
 \,,\nonumber\\
 \tilde H^{abcd}_3 &=& 
  \frac{1}{2}\left (f^{ace} f^{bde}  + f^{ade} f^{bce}\right) 
  \,,
\end{eqnarray}
one obtains
\begin{eqnarray}
  \lambda_1 &=& \tilde{\lambda_3} 
  \,,\nonumber\\
  \lambda_2 &=& \frac{1}{6} \left(\tilde{\lambda_1} + 2
  \tilde{\lambda_2}\right )
  \,,\nonumber\\
  \lambda_3 &=& \frac{1}{3}\left(- \tilde{\lambda_1} + 
  \tilde{\lambda_2}\right )
  \,.
\end{eqnarray}

The renormalization constants for the evanescent couplings
are defined through
\begin{align}
  g_e^0   &= \mu^{\epsilon}Z_e g_e\,,\qquad &
  \sqrt{\lambda_r^0} &= \mu^\epsilon Z_{\lambda_r} \sqrt{\lambda_r}\,,
  \qquad &
  \varepsilon^{0,a}_\sigma &= \sqrt{Z_3^\varepsilon} \varepsilon^a_\sigma\,,
  \nonumber\\
  \Gamma_{q\bar{q}\varepsilon}^{0} &= 
  Z_1^\varepsilon \Gamma_{q\bar{q}\varepsilon}\,,\qquad &
  \Gamma^{r,0}_{\varepsilon\varepsilon\varepsilon\varepsilon}
  &= Z_1^r \Gamma^{r}_{\varepsilon\varepsilon\varepsilon\varepsilon}
  \,,
  \label{eq::renconst}
\end{align}
where $\Gamma_{q\bar{q}\varepsilon}$  and
$\Gamma_{\varepsilon\varepsilon\varepsilon\varepsilon}$ are the
one-particle irreducible \epscalar{}--quark and
four-\epscalar{} Green functions, respectively, the superscript 
``0'' denotes bare quantities, and $\mu$ is the renormalization scale.
The charge renormalization constants are obtained from the following 
relations 
\begin{equation}
\begin{split}
  Z_e &= \frac{Z_1^\varepsilon}{Z_2\sqrt{Z_3^\varepsilon}} 
  \,,\qquad
  Z_{\lambda_r} = \frac{\sqrt{Z_1^r}}{Z_3^\varepsilon}
  \,,
\end{split}
\end{equation}
with $Z_2$ being the wave function renormalization constant of the quark
fields.

Let us introduce the couplings
\begin{eqnarray}
  \alpha_s = \frac{g_s^2}{4\pi}\,,
  \quad
  \alpha_e = \frac{g_e^2}{4\pi}
  \quad\mbox{and}\quad 
  \eta_r = \frac{\lambda_r}{4\pi}
  \,,
\end{eqnarray}
and define the corresponding $\beta$ functions in the \drbar{}
scheme:
\begin{eqnarray}
  \lefteqn{\betaDRbar_s(\asDRbar,\alpha_e,\{\eta_r\})
   = \mu^2 \frac{{\rm d}}{{\rm d}\mu^2} \apiDR}
   \nonumber\\
   &=& - \left[\epsilon \apiDR + 2 \frac{\asDRbar}{\ZsDRbar}
                 \left(
                 \frac{\partial \ZsDRbar}{\partial \alpha_e} \beta_e
                 + \sum_r \frac{\partial \ZsDRbar}{\partial \eta_r}
   \beta_{\eta_r} 
                 \right) \right]
   \left(1+ 2 \frac{\asDRbar}{\ZsDRbar} \frac{\partial \ZsDRbar}
   {\partial \asDRbar}\right)^{-1}    
   \nonumber\\
         &=& - \epsilon\apiDR - \sum_{i,j,k,l,m} \betaDRbar_{ijklm}
   \left(\apiDR\right)^i\left(\aepi\right)^j
   \left(\frac{\eta_1}{\pi}\right)^k
   \left(\frac{\eta_2}{\pi}\right)^l 
   \left(\frac{\eta_3}{\pi}\right)^m 
         \,,
    \label{eq::Zg_beta1}\\
   \lefteqn{\beta_e(\asDRbar,\alpha_e,\{\eta_r\})
                 = \mu^2 \frac{{\rm d}}{{\rm d}\mu^2} \aepi}
   \nonumber\\
   &=& - \left[\epsilon \aepi + 2 \frac{\alpha_e}{Z_e} 
                 \left(
                 \frac{\partial Z_e}{\partial \asDRbar} \betaDRbar_s
                 + \sum_r \frac{\partial Z_e}{\partial \eta_r} \beta_{\eta_r}
                 \right) \right]
   \left(1+ 2 \frac{\alpha_e}{Z_e} \frac{\partial Z_e}
   {\partial \alpha_e}\right)^{-1} 
   \nonumber\\
         &=& - \epsilon\aepi
    - \sum_{i,j,k,l,m} \beta^{e}_{ijklm} 
    \left(\apiDR\right)^i
    \left(\aepi\right)^j 
    \left(\frac{\eta_1}{\pi}\right)^k
    \left(\frac{\eta_2}{\pi}\right)^l 
    \left(\frac{\eta_3}{\pi}\right)^m 
    \,,
    \label{eq::Zg_beta2}\\
   \lefteqn{\beta_{\eta_r}(\asDRbar,\alpha_e,\{\eta_r\})
   = \mu^2 \frac{{\rm d}}{{\rm d}\mu^2} \frac{\eta_r}{\pi}}
   \nonumber\\
   &=&
   - \left[\epsilon \frac{\eta_r}{\pi}
     + 2 \frac{\eta_r}{Z_{\lambda_r}} 
     \left(
     \frac{\partial Z_{\lambda_r}}{\partial \asDRbar} \betaDRbar_s
     + \frac{\partial Z_{\lambda_r}}{\partial \alpha_e} \beta_e 
     + \sum_{r^\prime\not=r}
     \frac{\partial Z_{\lambda_r}}{\partial \eta_{r^\prime}}
     \beta_{\eta_{r^\prime}}  
     \right)\right]
   \left(1+ 2 \frac{\eta_r}{Z_{\lambda_r}} 
   \frac{\partial Z_{\lambda_r}}
        {\partial \eta_r}\right)^{-1}
                \nonumber\\
                &=& - \epsilon\frac{\eta_r}{\pi}
    - \sum_{i,j,k,l,m} \beta^{\eta_r}_{ijklm} 
    \left(\apiDR\right)^i
    \left(\aepi\right)^j 
    \left(\frac{\eta_1}{\pi}\right)^k
    \left(\frac{\eta_2}{\pi}\right)^l 
    \left(\frac{\eta_3}{\pi}\right)^m 
                \,.
    \label{eq::Zg_beta3}
\end{eqnarray}
Here and in the following we do not explicitely display the dependence
on the renormalization scale $\mu$, i.e., $\alpha_s\equiv \alpha_s(\mu)$
etc.  Note that in the second line of Eq.\,(\ref{eq::Zg_beta1}), the
${\cal O}(\epsilon)$ terms of $\beta_e$ and $\beta_{\eta_r}$ contribute
to the finite part of $\betaDRbar_s$, and similarly for
Eqs.\,(\ref{eq::Zg_beta2}) and (\ref{eq::Zg_beta3}).  As we will see
below, in order to compute the four-loop term of $\betaDRbar$ one needs
$\beta_e$ to two-loop and $\beta_{\eta_r}$ ($r=1,2,3$) to one-loop
order.

In analogy to the $\beta$ functions we introduce the quark mass
anomalous dimension which is defined through
\begin{eqnarray}
  \gammaDRbar_m(\asDRbar,\alpha_e,\{\eta_r\})
  &=& \frac{\mu^2}{\mDRbar} \frac{{\rm d}}{{\rm d}\mu^2} \mDRbar
  \nonumber\\
  &=& - \pi \betaDRbar_s
  \frac{\partial \ln \ZmDRbar}{\partial \asDRbar}
  -  \pi \beta_e \frac{\partial  \ln \ZmDRbar }{\partial
  \alpha_e}
  -\pi \sum_r \beta_{\eta_r} \frac{\partial  \ln \ZmDRbar
  }{\partial \eta_r}
        \nonumber\\
        &=& - \sum_{i,j,k,l,m} \gammaDRbar_{ijklm} 
  \left(\apiDR\right)^i
  \left(\aepi\right)^j 
  \left(\frac{\eta_1}{\pi}\right)^k
  \left(\frac{\eta_2}{\pi}\right)^l 
  \left(\frac{\eta_3}{\pi}\right)^m 
  \,.
  \label{eq::Zm_gamma}
\end{eqnarray}
From this equation one can see that for the four-loop term of
$\gammaDRbar_m$, the beta functions $\beta_e$ and $\beta_{\eta_r}$ are
needed to three- and one-loop order, respectively,
since the dependence of $\ZmDRbar$ on $\alpha_e$ ($\eta_r$) starts at
one-loop (three-loop) order~\cite{Harlander:2006rj}.

The one-loop terms for $\beta_{\eta_r}$ and the three-loop expression
for $\beta_e$ can be computed using standard techniques (see
e.g.~\reference{Harlander:1998dq}), leading to the following
non-vanishing coefficients in Eqs.\,(\ref{eq::Zg_beta2}) and
(\ref{eq::Zg_beta3}):
\begin{eqnarray}
        \beta_{04000}^e
        &=&
        -\frac{55}{432}
        -\frac{91}{48} \zeta_3
        -\left( \frac{725}{1152} - \frac{17}{96}\zeta_3 \right) n_f
        +\frac{55}{768} n_f^2
        \,,\nonumber\\
        \beta_{13000}^e
        &=&
        \frac{2423}{1728}
        +\frac{5}{36} \zeta_3
        -\left( \frac{313}{288} + \frac{5}{24}\zeta_3 \right) n_f
        +\frac{9}{64} n_f^2
        \,,\nonumber\\
        \beta_{22000}^e
        &=&
        \frac{189157}{13824}
        -\frac{11}{16} \zeta_3
        -\left( \frac{35543}{9216} - \frac{73}{32}\zeta_3 \right) n_f
        +\frac{55}{768} n_f^2
        \,,\nonumber\\
        \beta_{31000}^e
        &=&
        \frac{4589}{512}
        +\left( \frac{1157}{6912} - \frac{5}{3}\zeta_3 \right) n_f
        -\frac{415}{5184} n_f^2
        \,,\nonumber\\
        \beta_{03100}^e
        &=&
        -\frac{9}{64}
        +\frac{243}{128} n_f
        \,,\qquad
        \beta_{03010}^e
        \,\,=\,\, \frac{5}{8} - \frac{45}{64} n_f
        \,,\nonumber\\
        \beta_{03001}^e
        &=&
        \frac{3}{32}
        -\frac{81}{64} n_f
        \,,\qquad
        \beta_{12100}^e
        \,\,=\,\, -\frac{219}{16}
        \,,\qquad
        \beta_{12010}^e
        \,\,=\,\, \frac{145}{48}
        \,,\nonumber\\
        \beta_{12001}^e
        &=& \frac{73}{8}
        \,,\qquad
        \beta_{02200}^e
        \,\,=\,\,
        \frac{1413}{512} 
        -\frac{729}{1024} n_f
        \,,\qquad
        \beta_{02020}^e
        \,\,=\,\,
        -\frac{115}{32}
        +\frac{135}{64} n_f
        \,,\nonumber\\
        \beta_{02002}^e
        &=&
        -\frac{161}{256}
        -\frac{567}{512} n_f
        \,,\qquad
        \beta_{02110}^e
        \,\,=\,\, \frac{75}{8}
        \,,\qquad
        \beta_{02101}^e
        \,\,=\,\,
        -\frac{471}{128}
        +\frac{243}{256} n_f
        \,,\nonumber\\
        \beta_{02011}^e
        &=&
        -\frac{85}{8}
        \,,\qquad
        \beta_{21100}^e
        \,\,=\,\, -\frac{1125}{1024}
        \,,\qquad
        \beta_{21010}^e
        \,\,=\,\, \frac{105}{128}
        \,,\qquad
        \beta_{21001}^e
        \,\,=\,\, \frac{615}{512}
        \,,\nonumber\\
        \beta_{11200}^e
        &=& 
        \frac{891}{128}
        \,,\qquad
        \beta_{11020}^e
        \,\,=\,\, - \frac{45}{4}
        \,,\qquad
        \beta_{11002}^e
        \,\,=\,\, \frac{693}{64}
        \,,\qquad
        \beta_{11101}^e
        \,\,=\,\, - \frac{297}{32}
        \,,\nonumber\\
        \beta_{01300}^e
        &=&
        -\frac{1701}{1024}
        \,,\qquad
        \beta_{01003}^e
        \,\,=\,\, \frac{63}{128}
        \,,\qquad
        \beta_{01210}^e
        \,\,=\,\, - \frac{405}{128}
        \,,\qquad
        \beta_{01201}^e
        \,\,=\,\, \frac{1701}{512}
        \,,\nonumber\\
        \beta_{01120}^e
        &=&
        \frac{135}{32}
        \,,\qquad
        \beta_{01021}^e
        \,\,=\,\, - \frac{315}{32}
        \,,\qquad
        \beta_{01102}^e
        \,\,=\,\, - \frac{81}{128}
        \,,\nonumber\\
        \beta_{01012}^e
        &=&
        -\frac{315}{32}
        \,,\qquad
        \beta_{01111}^e
        \,\,=\,\, \frac{135}{16}
        \,,
        \label{eq::beta_e}
\end{eqnarray}
\begin{eqnarray}
        \beta_{20000}^{\eta_1}
        &=& -\frac{3}{8}
        \,,\qquad
        \beta_{10100}^{\eta_1}
        \,\,=\,\, \frac{9}{2}
        \,,\qquad
        \beta_{02000}^{\eta_1}
        \,\,=\,\, \frac{1}{3} n_f
        \,,\qquad
        \beta_{01100}^{\eta_1}
        \,\,=\,\, -\frac{1}{2} n_f 
        \,,\nonumber\\
        \beta_{00200}^{\eta_1}
        &=& -\frac{11}{8}
        \,,\qquad
        \beta_{00110}^{\eta_1}
        \,\,=\,\, -2
        \,,\qquad
        \beta_{00101}^{\eta_1}
        \,\,=\,\, \frac{7}{2}
        \,,\nonumber\\
        \beta_{20000}^{\eta_2}
        &=& -\frac{9}{16}
        \,,\qquad
        \beta_{10010}^{\eta_2}
        \,\,=\,\, \frac{9}{2}
        \,,\qquad
        \beta_{02000}^{\eta_2}
        \,\,=\,\, \frac{1}{24} n_f
        \,,\qquad
        \beta_{01010}^{\eta_2}
        \,\,=\,\, - \frac{1}{2} n_f
        \,,\nonumber\\
        \beta_{00200}^{\eta_2}
        &=& \frac{3}{16}
        \,,\qquad
        \beta_{00110}^{\eta_2}
        \,\,=\,\, \frac{1}{2}
        \,,\qquad
        \beta_{00101}^{\eta_2}
        \,\,=\,\, - \frac{1}{2}
        \,,\nonumber\\
        \beta_{00020}^{\eta_2}
        &=& - \frac{32}{3}
        \,,\qquad
        \beta_{00011}^{\eta_2}
        \,\,=\,\, - \frac{7}{6}
        \,,\qquad
        \beta_{00002}^{\eta_2}
        \,\,=\,\, \frac{7}{12}
        \,,\nonumber\\
        \beta_{10001}^{\eta_3}
        &=& \frac{9}{2}
        \,,\qquad
        \beta_{01001}^{\eta_3}
        \,\,=\,\, - \frac{1}{2} n_f
        \,,\qquad
        \beta_{00110}^{\eta_3}
        \,\,=\,\, 2
        \,,\qquad
        \beta_{00101}^{\eta_3}
        \,\,=\,\, \frac{5}{2}
        \,,\nonumber\\
        \beta_{00020}^{\eta_3}
        &=& \frac{10}{3}
        \,,\qquad
        \beta_{00011}^{\eta_3}
        \,\,=\,\, - \frac{20}{3}
        \,,\qquad
        \beta_{00002}^{\eta_3}
        \,\,=\,\, - \frac{7}{6}
        \,,
  \label{eq::beta_etar}
\end{eqnarray}
where $n_f$ is the number of active quark flavours and $\zeta_3 =
\zeta(3) = 1.20206...$, where $\zeta$ is Riemann's zeta function.
The one- and two-loop result of $\beta_e$ can be found in
\reference{Harlander:2006rj}.

\section{\label{sec::susy3}$\beta_e^{\rm SYM}$ and 
  $\beta_s^{\rm SYM}$ to three loops}
In order to check our technical setup, we calculated the $\beta$
functions of the strong and the evanescent coupling constant for a
\sic{} Yang Mills ({\abbrev SYM}) theory.

Such a theory is obtained from massless \qcd{} when replacing the quarks
by the \susy{} partner of the gluon, the so-called gluino.  Technically,
this amounts to changing the colour-matrix for the fermion-gluon coupling
of \qcd{} from the fundamental to the adjoint representation of the
gauge group. In addition, closed fermion loops have to be multiplied by
an extra factor 1/2 in order to take into account the Majorana
character of the gluino.

\susy{} requires that the gluino-gluon coupling $\alpha_s$ equals the
gluino-\epscalar{} coupling $\alpha_e$ to all orders of perturbation
theory, and therefore $\beta_e^{\rm SYM}=\beta_s^{\rm SYM}$ for the
respective $\beta$ functions.  However, it was found in
\reference{Avdeev:1982np} that this equality is violated at the
three-loop level. The result 
was interpreted~\cite{Avdeev:1982np, Avdeev:1982xy}  
as a manifestation of
explicit \susy{} breaking by \dred, when employed in the component (as opposed 
to the superfield) formalism.

Using the approach described above, we re-calculated $\beta_s^{\rm SYM}$
and $\beta_e^{\rm SYM}$ through three-loop order. When we set $\alpha_e
= \alpha_s$, the result for $\beta_s^{\rm SYM}$ agrees with
\reference{Avdeev:1981ew} ($C_A$ is the Casimir of the adjoint
representation of the gauge group):
\begin{equation}
\begin{split}
\beta_s^{\rm SYM} = -\left(\api\right)^2\,\left[
  \frac{3}{4}C_A
  + \frac{3}{8}C_A^2\api
  + \frac{21}{64}C_A^3\left(\api\right)^2\right] + {\cal O}(\alpha_s^5)\,.
\end{split}
\end{equation}
However, in contrast to \reference{Avdeev:1982np}, we find that
\begin{equation}
\begin{split}
\beta_e^{\rm SYM}=\beta_s^{\rm SYM} + {\cal O}(\alpha_s^5)\,.
\end{split}
\end{equation}

We draw the following conclusions from this result: {\rm (i)}~The
expression quoted in \reference{Avdeev:1982np} for the three-loop
expression of $\beta_e$ is incorrect; considering the fact that this
calculation has been done almost 25 years ago, it is probably impossible
to trace the origin of the difference to our result; {\rm (ii)}~in a
\sic{} Yang Mills theory, the evanescent coupling constant $\alpha_e$
renormalises in the same way as the strong coupling constant $\alpha_s$
up to three-loop level, as required by \susy{}; {\rm (iii)}~the setup of
our calculation has passed a strong consistency check.

\section{$\betaDRbar$ and $\gammaDRbar_m$ to four loops}\label{sec::4loop}

The direct way to compute the $\beta_s$ and $\gamma_m$ function is based
on the evaluation of the corresponding renormalization constants. For
such a calculation one can exploit that the divergent
parts of a logarithmically divergent integral is independent of the
masses and external momenta. Thus the latter can be chosen in 
a convenient way (provided no infrared divergences are introduced).
Up to three loops this procedure is quite well established and
automated programs exist to perform such calculations (see, e.g.,
Refs.~\cite{Larin:1991fz,Steinhauser:2000ry}).
Also at four-loop order this approach is feasible, however,
technically quite challenging. Thus we decided to adopt the 
indirect method discussed in Refs.~\cite{Bern:2002zk,Harlander:2006rj}.
It is based on the following formul\ae{} relating the quantities in 
\dred{} and \dreg{}:
\begin{eqnarray}
  \betaDRbar_s
  &=& \betaMSbar_s
  \frac{\partial \asDRbar}{\partial \asMSbar} + 
  \beta_e \frac{\partial \asDRbar}{\partial \alpha_e} +
  \sum_r \beta_{\eta_r} \frac{\partial \asDRbar}{\partial \eta_r}
  \,,
  \nonumber\\
  \gammaDRbar_m &=&
  \gammaMSbar_m \frac{\partial \ln \mDRbar}{\partial \ln \mMSbar}
  + \frac{\pi \betaMSbar_s}{\mDRbar} 
  \frac{\partial \mDRbar}{\partial \asMSbar}
  + \frac{\pi \beta_e}{\mDRbar}   
  \frac{\partial \mDRbar}{\partial \alpha_e}
  + \sum_r \frac{\pi \beta_{\eta_r}}{\mDRbar}   
  \frac{\partial \mDRbar}{\partial \eta_r}
  \,.
  \label{eq::DRED-DREG}
\end{eqnarray}
Let us in the following briefly discuss the order in perturbation theory
up to which the individual building blocks are needed. Of course, the
\msbar{} quantities are needed to four-loop order; they can be found in
Refs.~\cite{vanRitbergen:1997va,Chetyrkin:1997dh,
Vermaseren:1997fq,Czakon:2004bu}.  The dependence of $\asDRbar$ and
$\mDRbar$ on $\alpha_e$ starts at two- and one-loop
order~\cite{Harlander:2006rj}, respectively. Thus, $\beta_e$ is needed
up to the three-loop level (cf. Eq.~(\ref{eq::beta_e})).  On the other
hand, both $\asDRbar$ and $\mDRbar$ depend on $\eta_r$ starting from
three loops and consequently only the one-loop term of $\beta_{\eta_r}$
enters in Eq.~(\ref{eq::DRED-DREG}). It is given in
Eq.~(\ref{eq::beta_etar}).

Two further new ingredients are needed for the four-loop analysis,
namely, the three-loop relations between $\asDRbar$ and $\asMSbar$ and
between $\mDRbar$ and $\mMSbar$. The two-loop results have already been
presented in \reference{Harlander:2006rj}. Parametrising the three-loop
terms by $\delta_\alpha^{(3)}$ and $\delta_m^{(3)}$, we have
\begin{eqnarray}
  \asDRbar &=& \asMSbar\left[1+\frac{\asMSbar}{\pi} \frac{1}{4}
  +\left(\frac{\asMSbar}{\pi}\right)^2
  \frac{11}{8} 
  - \frac{\asMSbar}{\pi} \aepi
  \frac{1}{12} n_f
  + \delta_\alpha^{(3)} + \ldots \right]
  \,,
  \nonumber\\
  \mDRbar &=& \mMSbar\Bigg[1 -\aepi\frac{1}{3} +
  \left(\apiMS\right)^2 \frac{11}{48} -\apiMS\aepi
  \frac{59}{72}
  \nonumber\\
  &&\mbox{} + \left(\aepi\right)^2 \left( \frac{1}{6}
      +  \frac{1}{48} n_f\right)   
      + \delta_m^{(3)} + \ldots 
  \Bigg]
  \,,
  \label{eq::asMS2DR_2}
\end{eqnarray}
where the dots denote higher orders in $\asMSbar$, $\alpha_e$, and
$\eta_r$.  $\delta_\alpha^{(3)}$ and $\delta_m^{(3)}$ are obtained from
the finite parts of three-loop diagrams (see
\reference{Harlander:2006rj} for details). We find
\begin{eqnarray}
  \delta_\alpha^{(3)} &=&
        \left(\frac{\asMSbar}{\pi}\right)^3 \left(
        \frac{3049}{384}
        - \frac{179}{864} n_f \right)
        \nonumber\\
        && + \frac{\left(\asMSbar\right)^2}{\pi^3} \left(
        - \eta_1 \frac{9}{256}
        + \eta_2 \frac{15}{32}
        + \eta_3 \frac{3}{128}
        - \alpha_e \frac{887}{1152} n_f
        \right)
        \nonumber\\
        && + \frac{\asMSbar}{\pi^3} \left[
        \eta_1^2 \frac{27}{256}
        - \eta_2^2 \frac{15}{16}
        - \eta_1 \eta_3 \frac{9}{64}
        + \eta_3^2 \frac{21}{128}
        + \alpha_e^2 \left(     \frac{43}{864} n_f 
+ \frac{19}{1152} n_f^2 \right)
        \right]
  \,,
  \nonumber\\
  \delta_m^{(3)} &=&
        \left(\apiMS\right)^3 \left( 
        \frac{2207}{864} + \frac{19}{648} n_f
        \right)
        - \frac{\left(\asMSbar\right)^2\alpha_e}{\pi^3} \left(
        \frac{62815}{20736} + \frac{253}{1728} n_f - \frac{25}{72} \zeta_3
        \right)
        \nonumber\\
        && + \frac{\asMSbar\alpha_e^2}{\pi^3} \left[
        \frac{1973}{2592} - \frac{5}{36} \zeta_3 + \left( 
        \frac{103}{1728} + \frac{5}{36} \zeta_3 \right) n_f 
        \right]
        \nonumber\\
        && -\left(\aepi\right)^3 \left(
        \frac{7}{144} + \frac{5}{216}\zeta_3 + \frac{31}{576} n_f -
        \frac{5}{576} n_f^2 \right)
        - \frac{\alpha_e^2\eta_2}{\pi^3} \frac{5}{24}
        \nonumber\\
        && -\frac{\alpha_e}{\pi^3} \left(
        \eta_1^2\frac{9}{256} - \eta_2^2\frac{15}{16} 
        - \eta_1\eta_3\frac{3}{64} + \eta_3^2\frac{7}{128}
        \right)
  \,.
  \label{eq::as_mq_MS2DR}
\end{eqnarray}
We performed the corresponding calculation for arbitrary gauge parameter
and use the independence of the \msbar{}--\drbar{} relation as a check
of our result.  Furthermore, let us stress that also the cancellation of
the explicit $\ln\mu^2$ terms which occur at intermediate steps of the
calculation is non-trivial.

Inserting Eq.~(\ref{eq::as_mq_MS2DR}) into Eq.~(\ref{eq::DRED-DREG}) gives
for the four-loop coefficients of the $\beta$ function
\begin{eqnarray}
  \betaDRbar_{50000} 
        &=& 
        \frac{\beta_3}{256}
        +\frac{166861}{6144}
        -\frac{9109}{6912} n_f
        +\frac{457}{20736} n_f^2
        \,,\qquad
        \betaDRbar_{41000}
        \,\,=\,\,
        -\frac{1667}{512} n_f
        +\frac{145}{2304} n_f^2
        \,,\nonumber\\
        \betaDRbar_{32000}
        &=&
        -\frac{409}{6912} n_f
        +\frac{1303}{4608} n_f^2
        \,,\qquad
        \betaDRbar_{23000}
        \,\,=\,\,
        \frac{5}{1296} n_f
        -\frac{49}{3456} n_f^2
        -\frac{19}{2304} n_f^3
        \,,\nonumber\\
        \betaDRbar_{40100}
        &=&
        -\frac{171}{512}
        +\frac{3}{512} n_f
        \,,\qquad
        \betaDRbar_{40010}
        \,\,=\,\,
        \frac{285}{64}
        -\frac{5}{64} n_f
        \,,\qquad
        \betaDRbar_{40001}
        \,\,=\,\,
        \frac{57}{256}
        -\frac{1}{256} n_f
        \,,\nonumber\\
        \betaDRbar_{31100} &=& \frac{9}{512} n_f
        \,,\qquad
        \betaDRbar_{31010} \,\,=\,\, -\frac{15}{64} n_f
        \,,\qquad
        \betaDRbar_{31001} \,\,=\,\, -\frac{3}{256} n_f
        \,,\nonumber\\
        \betaDRbar_{30200} &=& \frac{2223}{2048}
        \,,\qquad
        \betaDRbar_{30020} \,\,=\,\, -\frac{855}{64}
        \,,\qquad
        \betaDRbar_{30002} \,\,=\,\, \frac{441}{256}
        \,,\qquad
        \betaDRbar_{30110} \,\,=\,\, \frac{45}{128}
        \,,\nonumber\\
        \betaDRbar_{30101} &=& -\frac{801}{512}
        \,,\qquad
        \betaDRbar_{30011} \,\,=\,\, -\frac{45}{64}
        \,,\qquad
        \betaDRbar_{22100} \,\,=\,\, \frac{21}{128} n_f
        \,,\qquad
        \betaDRbar_{22010} \,\,=\,\, -\frac{35}{192} n_f
        \,,\nonumber\\
        \betaDRbar_{22001} &=& -\frac{7}{64} n_f
        \,,\qquad
        \betaDRbar_{21200} \,\,=\,\, -\frac{9}{64} n_f
        \,,\qquad
        \betaDRbar_{21020} \,\,=\,\, \frac{5}{4} n_f
        \,,\qquad
        \betaDRbar_{21002} \,\,=\,\, -\frac{7}{32} n_f
        \,,\nonumber\\
        \betaDRbar_{21101} &=& \frac{3}{16} n_f
        \,,\qquad
        \betaDRbar_{20300} \,\,=\,\, -\frac{297}{1024}
        \,,\qquad
        \betaDRbar_{20030} \,\,=\,\, 20
        \,,\qquad
        \betaDRbar_{20003} \,\,=\,\, -\frac{49}{128}
        \,,\nonumber\\
        \betaDRbar_{20210} &=& -\frac{135}{128}
        \,,\qquad
        \betaDRbar_{20201} \,\,=\,\, \frac{297}{512}
        \,,\qquad
        \betaDRbar_{20120} \,\,=\,\, -\frac{45}{32}
        \,,\qquad
        \betaDRbar_{20021} \,\,=\,\, \frac{105}{32}
        \,,\nonumber\\
        \betaDRbar_{20102} &=& \frac{63}{128}
        \,,\qquad
        \betaDRbar_{20012} \,\,=\,\, -\frac{105}{32}
        \,,\qquad
        \betaDRbar_{20111} \,\,=\,\, \frac{45}{16}
        \,,
\end{eqnarray}
where the four-loop \msbar{} coefficient $\beta_3$ is given in
Eq.~(8) of \reference{vanRitbergen:1997va}.

Similarly we obtain for the 
four-loop coefficient of the mass anomalous dimension
\begin{eqnarray}
  \gammaDRbar_{40000} &=& 
  \gamma_3 
        - \frac{18763}{2304}
        +\left( \frac{1}{6} + \frac{5}{8}\zeta_3 \right) n_f
        +\frac{29}{5184} n_f^2
  \,,\nonumber\\
        \gammaDRbar_{31000}     &=&
        -\frac{147659}{4608} + \frac{125}{48}\zeta_3
        +\left( \frac{58253}{31104} + \frac{95}{216}\zeta_3 \right) n_f
        +\frac{407}{7776} n_f^2
        \,,\nonumber\\
        \gammaDRbar_{22000} &=&
        -\frac{134147}{62208} - \frac{281}{432}\zeta_3
        +\left( \frac{336497}{124416} + \frac{49}{432}\zeta_3 \right) n_f
        -\left( \frac{181}{10368} + \frac{5}{216}\zeta_3 \right) n_f^2
        \,,\nonumber\\
        \gammaDRbar_{13000} &=&
        -\frac{595}{7776} -\frac{25}{108}\zeta_3
        -\left( \frac{1163}{10368} - \frac{5}{27}\zeta_3 \right) n_f
        -\left( \frac{145}{3456} + \frac{5}{72}\zeta_3 \right) n_f^2
        \,,\nonumber\\
        \gammaDRbar_{04000} &=&
        \frac{191}{2592} + \frac{67}{108}\zeta_3
        +\left( \frac{301}{1728} - \frac{1}{24}\zeta_3 \right) n_f
        +\frac{5}{384} n_f^2
        -\frac{5}{768} n_f^3
        \,,\nonumber\\
        \gammaDRbar_{30100} &=& \frac{9}{256}
        \,,\qquad
        \gammaDRbar_{30010} \,\,=\,\, -\frac{15}{32}
        \,,\qquad
        \gammaDRbar_{30001} \,\,=\,\, -\frac{3}{128}
        \,,\qquad
        \gammaDRbar_{21100} \,\,=\,\, \frac{201}{512}
        \,,\nonumber\\
        \gammaDRbar_{21010} &=& -\frac{85}{64}
        \,,\qquad
        \gammaDRbar_{21001} \,\,=\,\, -\frac{107}{256}
        \,,\qquad
        \gammaDRbar_{20200} \,\,=\,\, -\frac{27}{256}
        \,,\qquad
        \gammaDRbar_{20020} \,\,=\,\, \frac{15}{16}
        \,,\nonumber\\
        \gammaDRbar_{20002} &=& -\frac{21}{128}
        \,,\qquad
        \gammaDRbar_{20101} \,\,=\,\, \frac{9}{64}
        \,,\qquad
        \gammaDRbar_{12100} \,\,=\,\, \frac{351}{64}
        \,,\qquad
        \gammaDRbar_{12010} \,\,=\,\, -\frac{365}{96}
        \,,\nonumber\\
        \gammaDRbar_{12001} &=& -\frac{117}{32}
        \,,\qquad
        \gammaDRbar_{11200} \,\,=\,\, -\frac{1563}{512}
        \,,\qquad
        \gammaDRbar_{11020} \,\,=\,\, \frac{1645}{96}
        \,,\qquad
        \gammaDRbar_{11002} \,\,=\,\, -\frac{3647}{768}
        \,,\nonumber\\
        \gammaDRbar_{11101} &=& \frac{521}{128}
        \,,\qquad
        \gammaDRbar_{03100} \,\,=\,\, 
        -\frac{13}{64} 
        -\frac{45}{64} n_f
        \,,\qquad
        \gammaDRbar_{03010} \,\,=\,\,
        \frac{55}{96} n_f
        \,,\nonumber\\
        \gammaDRbar_{03001} &=&
        \frac{13}{96}
        +\frac{15}{32} n_f
        \,,\qquad
        \gammaDRbar_{02200} \,\,=\,\,
        -\frac{223}{256}
        +\frac{153}{512} n_f
        \,,\qquad
        \gammaDRbar_{02020} \,\,=\,\,
        \frac{395}{144}
        -\frac{65}{32} n_f
        \,,\nonumber\\
        \gammaDRbar_{02002} &=&
        \frac{259}{1152}
        +\frac{119}{256} n_f
        \,,\qquad
        \gammaDRbar_{02110} \,\,=\,\, -\frac{155}{48}
        \,,\qquad
        \gammaDRbar_{02101} \,\,=\,\,
        \frac{233}{192}
        -\frac{51}{128} n_f
        \,,\nonumber\\
        \gammaDRbar_{02011} &=& \frac{545}{144}
        \,,\qquad
        \gammaDRbar_{01300} \,\,=\,\, \frac{333}{512}
        \,,\qquad
        \gammaDRbar_{01030} \,\,=\,\, -20
        \,,\qquad
        \gammaDRbar_{01003} \,\,=\,\, -\frac{7}{192}
        \,,\nonumber\\
        \gammaDRbar_{01210} &=& \frac{105}{64}
        \,,\qquad
        \gammaDRbar_{01201} \,\,=\,\, -\frac{333}{256}
        \,,\qquad
        \gammaDRbar_{01120} \,\,=\,\, -\frac{5}{16}
        \,,\qquad
        \gammaDRbar_{01021} \,\,=\,\, \frac{35}{48}
        \,,\nonumber\\
        \gammaDRbar_{01102} &=& \frac{3}{64}
        \,,\qquad
        \gammaDRbar_{01012} \,\,=\,\, \frac{245}{48}
        \,,\qquad
        \gammaDRbar_{01111} \,\,=\,\, -\frac{35}{8}
        \,,
\end{eqnarray}
where the four-loop \msbar{} coefficient $\gamma_3$ can be found in
Eq.~(6) of~\reference{Chetyrkin:1997dh}.

\section{\label{sec::moresusy}The four-loop supersymmetric case}

We saw in Section~\ref{sec::susy3} that in the special case of \susy,
the relation $\alpha_s = \alpha_e$ is preserved by the $\beta$-functions
through three loops. We now consider the \sic{} case at the four loop
level.  The results in Section~\ref{sec::4loop} were presented for the
gauge group $SU(3)$; however, we have evaluated those parts that are not
related to the evanescent couplings $\eta_2$ and $\eta_3$ also for a general 
Lie group $G$. It is 
well-known that a simple change of color factors,\footnote{Note that
this procedure differs from the one outlined in
Section~\ref{sec::susy3}, where we modified the Feynman rules in color
space and re-evaluated the color factor for each diagram. Here, we
simply replace $C_F$ (fundamental Casimir) and $T$ (fundamental trace
normalization) by $C_A$ (adjoint Casimir) in the final result.}  in
addition to the statistical factor 1/2 for closed fermion loops,
translates these terms into a \sic{} Yang Mills theory. In this way, we
can compare our four-loop results to the gauge $\beta$-function
$\beta_s^{\rm SYM}$ which was presented in 1998~\cite{Jack:1998uj}:
\begin{equation}
\begin{split}
\beta_s^{\rm SYM} = -\left(\api\right)^2\,\left[
  \frac{3}{4}C_A
  + \frac{3}{8}C_A^2\api
  + \frac{21}{64}C_A^3\left(\api\right)^2
  + \frac{51}{128}C_A^4\left(\api\right)^3
\right] + {\cal O}(\alpha_s^6)\,.
\end{split}
 \label{eq::betag4}
\end{equation}
The method employed in \reference{Jack:1998uj} to obtain the four-loop result 
was very indirect, in particular relying on the \nsvz{} 
form~\cite{Jones:1983ip, Novikov:1983ee} of $\beta_s^{\rm SYM}$.
It is therefore a remarkable check on our calculations, and indeed those of 
\reference{Jack:1998uj}, that we obtain 
precise agreement with Eq.~(\ref{eq::betag4})
when we adapt our calculation to the \sic\ case, as described above. 
(As well as setting $\alpha_e = \alpha_s$ 
we must of course set $\eta_1 = \alpha_s, \eta_2 = \eta_3 = 0$). 

Turning to the mass anomalous dimension we have a similar powerful check. 
In the \sic\ case the fermion mass term breaks \susy; 
but  $\gamma_m$ (a.k.a.\ the gaugino $\beta$-function) 
is in fact given in terms of $\beta_s$ by 
the simple equation~\cite{Jack:1997pa}:
\begin{equation}
\gamma_m^{\rm SYM} = \pi \alpha_s\frac{\rm d}{\rm d\alpha_s}
\left[\frac{\beta_s^{\rm SYM}}{\alpha_s}\right].
\label{eq::exactgamma}
\end{equation}
(This relationship\footnote{Note that our definition of $\gamma_m$ 
in Eq.~(\ref{eq::Zm_gamma}) differs by a factor of two 
(and a factor of $M$) from the definition of $\beta_M$ in 
\reference{Jack:1997pa}.}
between $\gamma_m$ and $\beta$ holds in both 
\dred{}  and \nsvz{} schemes.) 
Through four loops we have at once from Eq.~(\ref{eq::exactgamma}) that 
\begin{equation}
\begin{split}
\gamma_m^{\rm SYM} = -\left(\api\right)\,\left[
  \frac{3}{4}C_A
  + \frac{3}{4}C_A^2\api
  + \frac{63}{64}C_A^3\left(\api\right)^2
  + \frac{51}{32}C_A^4\left(\api\right)^3
\right] + {\cal O}(\alpha_s^6)\,.
\end{split}
 \label{eq::gamma4}
\end{equation}
Quite remarkably, in the \sic\ case we again find this agrees with our 
calculations. This is again striking confirmation of our methodology 
and of the exact result Eq.~(\ref{eq::exactgamma}).

\section{\label{sec::concl}Conclusions}

In this paper we have applied \dred{} to \qcd, and calculated 
both the gauge $\beta$-function and the mass anomalous dimension 
to the four-loop level. These calculations required careful 
treatment of the evanescent Yukawa and quartic couplings of the 
\epscalar{}. In the \sic{} limit we explicitly verified that 
the $\beta$-function for the evanescent Yukawa coupling reproduces the 
gauge $\beta$-function through three loops. 

The popularity of the \mssm{} and the construction of the {\abbrev CERN}
Large Hardon Collider ({\abbrev LHC}) has led to many increasingly
precise calculations of sparticle production and decay processes, using
\dred{}.
The \mssm{} is a softly broken {\it \sic\/} theory, so  we might well
expect its dimensionless coupling sector to renormalise like the
underlying  \sic\ theory, without worrying about evanescent couplings;
to test this (in the same manner as described above) 
will  require a generalisation  of our calculations to
incorporate scalar fields. In the \mssm{}, however, 
there is in fact one evanescent quantity which 
must certainly be considered: the
\epscalar{} mass~\cite{jj}. This exists also in \qcd, but affects
neither the gauge $\beta$-function nor the fermion mass anomalous
dimension on simple dimensional grounds so we have not considered it
here.

If, however, one wants to match \mssm{} calculations on to the Standard Model 
(or, for example,  consider an intermediate energy theory
produced  by integrating out the squarks and
sleptons~\cite{Arkani-Hamed:2004fb}) then evidently the use of \dred{}
inevitably means one must worry about  evanescent couplings. 
\reference{Harlander:2006rj} pointed out a couple of instances where 
naive treatment of the evanescent couplings has led to 
incorrect conclusions; we believe that careful treatment of the 
\epscalar{}s in higher order calculations 
as presented in \reference{Harlander:2006rj}
and here will prove invaluable in matching \mssm{} 
calculations to low energy physics.

\vspace*{1em}

\noindent
{\large\bf Acknowledgements}\\ 
One of us (DRTJ) thanks Ian Jack for conversations, and 
both CERN and KITP (Santa Barbara) for hospitality. This work was 
supported by the DFG through SFB/TR~9 and the {\it
Emmy Noether} program HA\,2990/2-1, and by 
the National Science Foundation  
under Grant No. PHY99-07949.

\end{document}